\documentclass[journal]{IEEEtran}
\usepackage{epsfig}
\usepackage{graphicx}
\usepackage{color}
\usepackage{subfigure}

\usepackage[para]{threeparttable}
\usepackage{amssymb}
\usepackage{amsmath}
\usepackage{graphicx}
\usepackage{epsf}
\usepackage{epsfig}
\usepackage{psfig}
\usepackage{ccaption}
\usepackage{array}
\usepackage{tabularx}
\usepackage{multirow}
\usepackage{epsfig}
\usepackage{cite}
\usepackage{procedure,algorithm,algorithmic}

\begin{document}
\twocolumn

%
\title{Time Synchronization Attack in Smart Grid-\\Part II: Cross Layer Detection Mechanism}

\author{Zhenghao Zhang,
        Matthew Trinkle,
        Aleksandar D. Dimitrovski,
        and Husheng Li,
\thanks{Z. Zhang, and H. Li are with Department
of Electrical Engineering and Computer Science, University of Tennessee, Knoxville, TN.
M. Trinkle is with Sensor Signal Processing Group, University of Adelaide, Australia.
A. D. Dimitrovski is with Energy and Transportation Sciences Division,
Oak Ridge National Lab, Oak Ridge, TN.
The research is under the support of National Science Foundation under
grants  ECCS-0901425.
}}

\markboth{{\em This paper has been submitted to IEEE Transaction on Smart Grid}}%
{Zhang \MakeLowercase{\textit{et al.}}: IEEE Transaction on Smart Grid}

\maketitle

\begin{abstract}
A novel time synchronization attack (TSA) on wide area monitoring systems in smart grid has been identified in the first part of this paper.
A cross layer detection mechanism is proposed to combat TSA in part II of this paper. In the physical layer, we propose a GPS carrier signal noise ratio (C/No) based spoofing detection technique. In addition, a patch-monopole hybrid antenna is applied to receive GPS signal. By computing the standard deviation of the C/No difference from two GPS receivers, a priori probability of spoofing detection
is fed to the upper layer, where power system state is estimated and controlled. A trustworthiness based evaluation method is applied to identify the PMU being under TSA. Both the physical layer and upper layer algorithms are integrated to detect the TSA, thus forming a cross layer mechanism.
Experiment is carried out to verify the effectiveness of the proposed TSA detection algorithm.
\end{abstract}


\IEEEpeerreviewmaketitle

\section{Introduction}
The security of smart grid has become an important research topic \cite{Grid_future}\cite{SCADA}\cite{DataAttack_Xie},
since the electricity system is closely related to many critically important aspects of modern society.
The secure wide area monitoring system (WAMS) \cite{FNET_SGTRAN} has become the key
component in maintaining the reliability of the entire power system.
As a complex subsystem of the smart gird, WAMS has faced many challenges on its security due to
its widely distributed monitoring devices and extensive communication infrastructure \cite{Khurana_SecurSG}.

In part I of this paper, we identified a novel potential attack to WAMS in smart grid,
namely the time synchronization attack (TSA) through GPS spoofing. Furthermore, we have analyzed and demonstrated the impact of TSA on three
different WAMS applications, which showed that TSA can confuse the control center with a wrong system operation status
by introducing counterfeit time stamps to the true measurements. Since the malicious attackers do not need to physically
connect to the monitoring device or the communication network, TSA cannot be prevented by simply enhancing the firmware of the
monitoring devices. In addition, unlike a malicious data attack \cite{DataAttack_Kosut}\cite{BadDataDetection_Lin},
TSA does not change the monitoring data. Therefore, common defense methods for malicious data attacks are not suitable to combat TSA.

Motivated by the extreme importance of the security of WAMS in smart grid,
in part II of this paper, we study the detection of TSA to ensure the reliability
of the monitoring system. In the physical layer of WAMS, TSA can be mitigated by
techniques of GPS spoofing detection, which can be implemented in each individual GPS receiver \cite{GPSAntispoof_Shepard}.
These techniques are based on the GPS signal parameters obtained directly from the GPS receiver \cite{GPSAntispoof_Ledvina}, such
as its position solution \cite{GPSAntispoof_Daneshmand}\cite{GPSAntispoof_Nielsen}\cite{GPSAntispoof_Wen},
the Doppler shift of the GPS satellites \cite{GPSAntispoof_Cavaleri2}\cite{GPSAntispoof_Papadimitratos},
or the SNR of the GPS signals \cite{GPSAntispoof_Cavaleri1}.

An effective low-cost implementation of an anti-spoofing technique can simply compare the position solutions from
two GPS receivers close to each other, since a spoofing signal would cause both receivers to report the same position \cite{GPSAntispoof_Wen}.
To that end, the receivers need to be sufficiently far away from each other such that the two position solutions are easily separated in the absence of a spoofer; meanwhile,
they should also be close enough such that they are affected by the same spoofing signal. For closely spaced receivers, the phase of the GPS signals at each
receiver antenna can be used to detect the presence of a spoofer. As the spoofer signal comes from a particular direction, the phase shift between the two
antennas will be identical for all satellites, for the spoofing signal, which is not true for real GPS signals coming from different
directions.  This technique can be further improved by predicting the expected phase shifts from the satellite orbit models.  However, this technique
requires the GPS receiver to measure and report the GPS carrier phase measurement, which may not be valid in practice.
Angle-of-arrival (AOA) based spoofing detection (AOASD) \cite{GPSAntispoof_Montgomery} has been considered to be among the most effective techniques to
detect a GPS spoofing attack. Typically, a GPS receiver receives navigation signal from multiple GPS satellites which have different AOAs.
In a sharp contrast, counterfeit navigation signals from different GPS satellites have the same AOA,
since all the GPS signal is transmitted by one GPS spoofer. However, the AOA based techniques typically require an antenna
array with dedicated GPS receiver electronics to estimate the AOA of the GPS signals, which significantly increases the size and cost of device.

\begin{figure}[htcp]
  \centering
  \includegraphics[scale=0.7]{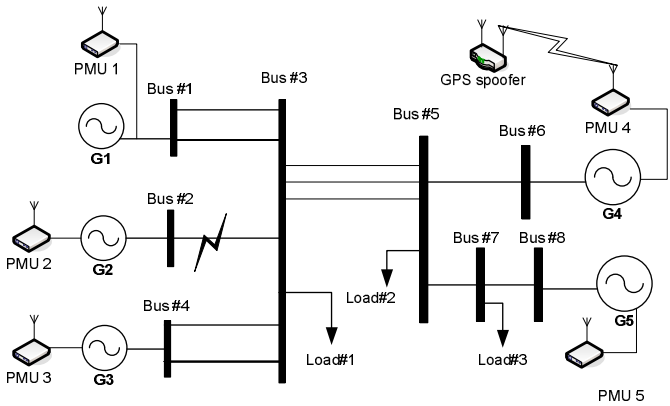}
  \caption{Illustration of trustiness evaluation based on power network modeling}\label{Fig:FiveMachine}
\end{figure}

In this paper, we propose a cross layer detection mechanism to protect WAMS against TSA.
In the physical layer, we propose a Carrier to Noise Measurements (C/No) based spoofing detection
algorithm. The difference of the C/No obtained from two GPS receivers is analyzed to detect the TSA.
In particular, we propose a monopole-patch hybrid antenna to detect a spoofing attack. The monopole
antenna is connected to one GPS receiver and the patch antenna is connected to the other.
Since the monopole and patch antenna have different radiation patterns in elevation,
the difference of the C/No measurements between the two receivers will vary with the elevation angle of the signal.
The proposed system can thus discriminate the elevation AOA between signals without requiring an expensive multi-element antenna array.
This technique is also expected to be effective against a cooperative GPS spoofing attack, as all the
GPS spoofing signals are likely to come from the same elevation angle near the horizon.
Furthermore, the output of the C/No based spoofing detection technique provides a probability indicating
whether the individual GPS receiver is experiencing spoofing.
Besides the physical layer TSA detection, we also propose a
trustworthiness framework based TSA detection in the upper layer, based on Kalman filtering and cross-check. Figure \ref{Fig:FiveMachine} provides an illustration of
how the trustworthiness of a set of
monitoring devices could be evaluated \cite{Trust_Li}.

As illustrated in Figure \ref{Fig:FiveMachine}, the entire power network consists of five generators (G1, G2, G3, G4, G5), eight buses and three loads.
There are five phasor measurement units (PMUs) distributed in the power network serving as the monitoring devices \cite{PMU_Dalaree}.
Assume that PMU-4 is suffering TSA from a malicious attacker and that a transmission line fault occurs between bus2 and bus3, which causes a disturbance in the power network system.
The distributed PMUs capture the dynamic feature of the system and report to the control center. Based on the time stamps
attached to the measurements sampled by the PMUs, the control center aligns the measurements from different PMUs to infer the location of the fault.
However, the measurements from PMU-4 cannot be aligned with those from other PMUs due to TSA. The measurements from different PMUs are correlated
according to the interconnection of the entire network. Therefore, the trustworthiness of each PMU
can be evaluated based on the system model and the system states. In this paper, we further combine both detection schemes in the physical and upper layers, thus forming a cross layer detection of TSA.

The remainder of this paper is organized as follows.
Section \ref{sec: PhyLayer} provides the physical layer TSA detection algorithm.
Section \ref{sec: Trustiness} presents the trustworthiness evaluation mechanism based
on the interconnection modeling of the power grid.
The cross layer TSA detection algorithm is given in Section \ref{sec: CrossLayer}.
Conclusions and future work are provided in Section \ref{sec:Conclusion}.


\section{TSA Detection in Physical Layer}\label{sec: PhyLayer}
In this section we propose a simple spoofer detection algorithm based on the Carrier to Noise Measurements (C/No) \cite{Book_GPSreceiver} obtained from two closely spaced GPS
receivers. This technique could be implemented in an existing system by simply logging the C/No measurements of the existing
GPS receiver and adding one more low cost GPS receiver module.

\subsection{C/No based spoofing detection}
This method uses the (C/No) measurement from two GPS receivers connected to two antennas with different radiation
patterns $G_1(\theta,\phi)$ and $G_2(\theta,\phi)$ to implement a mono-pulse system where the power ratio between two antennas is defined as
\begin{equation}\label{Eq.R_i}
    R_i=10\log_{10}(\frac{G_1(\theta_i,\phi_i)}{G_2(\theta_i,\phi_i)}).
\end{equation}
$R_i$ indicates the direction of arrival of the $i^{th}$ GPS signal arriving from azimuth direction $\theta_i$ and elevation direction $\phi_i$.
For the spoofing signal, all GPS signals are coming from the same direction as $\theta_1=\theta_2=...=\theta_I$ and hence they should have the
same power ratio, $R_i$ between the two antennas.
For the actual GPS signals $R_i$ should differ for each satellite as the signals come from different directions.  As a result, the standard
deviation of $R_i$ for all satellites observed at a given time point will be used to determine the likelihood of a spoofing signal. This technique
could also make use of other observables such as the satellite Doppler shift and the final position solution to improve robustness.

The C/No based spoofing detection works as follows:
At each time point, the value of $R_i$ for all observable GPS satellites is estimated from the C/No values obtained from the GPS receiver according to
\begin{equation}\label{Eq.RNCN}
    R_i = (C/No)_{i,1}(dB)-(C/No)_{i,2}(dB),
\end{equation}
where $(C/No)_(i,m)$ is the carrier to noise ratio of the $i^{th}$ GPS signal from the $m^{th}$ antenna in dB.
The standard deviation of $R_i$ is then compared with a threshold to determine the presence of a spoofing signal.  The threshold is calculated from
the two probability density functions (PDFs) of the standard deviation of the $R_i$ values when a spoofer is present and absent to achieve a given
false alarm rate. The two PDF's of the standard deviation of $R_i$ for a signal present and absent were obtained by conducting a field experiment.

\subsection{Field experiment for PDF}\label{subsec:Experiment}
\begin{figure}[]
\vspace{0pt}
\subfigure[C/No based detection experiment setup\label{Fig:Setup}]{
\begin{minipage}[b]{1\linewidth}
\centering
\includegraphics[scale=0.4]{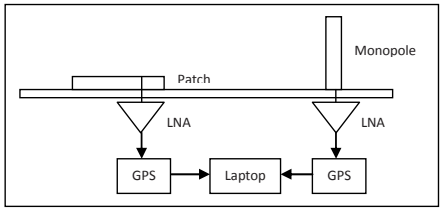}
\end{minipage}}\\%
\subfigure[Monopole-patch hybrid antenna\label{Fig:Antenna}]{
\begin{minipage}[b]{1\linewidth}
\centering
\includegraphics[scale=0.5]{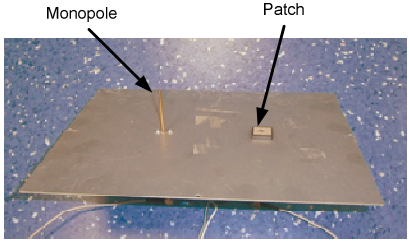}
\end{minipage}}
\caption{Basic setup for spoofing experiment.} \label{Fig:CNDetect}
\end{figure}

An experiment was set up using two GPS receivers, one connected to a patch antennas and one to a monopole antenna.
The C/No values of each receiver along with the satellite elevation and azimuth angle were logged. The antenna gain patterns are significantly
different as the radiation pattern of the patch has a maximum at 90 degrees elevation, while the monopole has a minimum at this angle.
The basic setup for the spoofing experiment is shown in Figure \ref{Fig:CNDetect}.

\begin{figure}[htpb]
  \centering
  \includegraphics[scale=0.4]{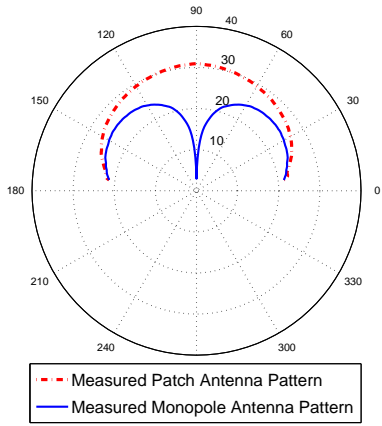}
  \caption{Radiation patterns of the monopole and patch antenna}\label{fig:Antenna}
\end{figure}

The radiation patterns of the patch and monopole were measured in an anechoic chamber and are shown in Figure \ref{fig:Antenna}.
Note that the antenna patterns are expected to be relatively uniform in azimuth, thus only the elevation cut is shown.  The monopole
antenna has reasonably good gain at all low elevation angles where it is most likely that a spoofer will attack.
The C/No measurements from both receivers were logged for approximately $90$ minutes
at two different times of the day.
These measurements were then used to build up a histogram of the C/No power ratios, $R_i$, which were used to generated an estimated
probability distribution (PDF) of the standard deviation of $R_i$ for the GPS signals with no spoofer present.

\begin{figure}[]
\vspace{0pt}
\subfigure[Laboratory setup diagram\label{Fig:Setup}]{
\begin{minipage}[b]{1\linewidth}
\centering
\includegraphics[scale=0.5]{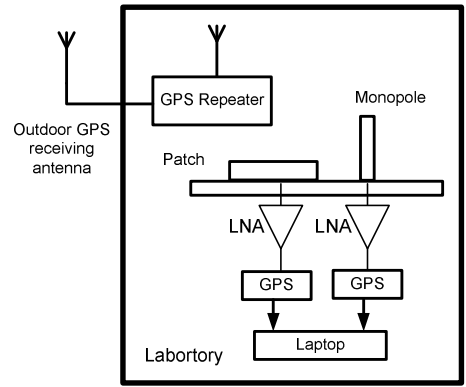}
\end{minipage}}\\%
\subfigure[Photo of the laboratory and devices\label{Fig:Antenna}]{
\begin{minipage}[b]{1\linewidth}
\centering
\includegraphics[scale=0.6]{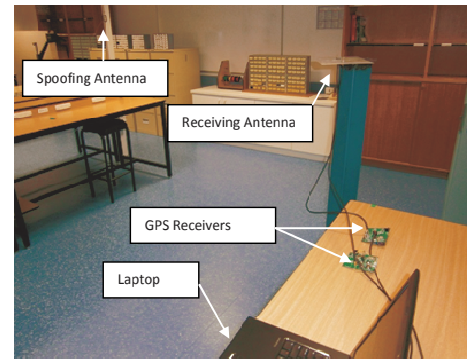}
\end{minipage}}
\caption{Illustration of the experiment setup.} \label{Fig:Exsetup}
\end{figure}

A PDF of the standard deviation of the $R_i$  values observed from a spoofing signal was also obtained by setting up the antenna array in the Laboratory where the
 GPS receiver was unable to track most of the real GPS signals, and then introducing the spoofing signal by using a GPS repeater. The GPS repeater was used to re-radiate the GPS
signals inside the Laboratory from a single antenna element, thus simulating a real spoofing environment. The C/No values from the GPS receivers connected to both
antennas were logged to estimate the PDF of the standard deviation of the $R_i$. The basic setup for the spoofing experiment is shown in Figure \ref{Fig:Exsetup}

\begin{figure}[htpb]
  \centering
  \includegraphics[scale=0.4]{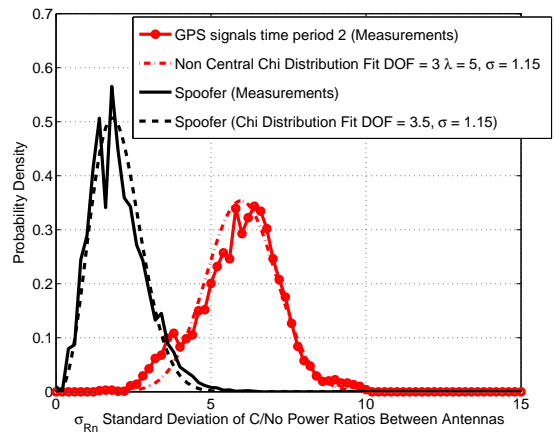}
  \caption{Estimated PDF of the standard deviation of the power ratio Rn between antennas for real GPS signals and spoofer}\label{fig:PDF}
\end{figure}

\begin{figure}[htpb]
  \centering
  \includegraphics[scale=0.4]{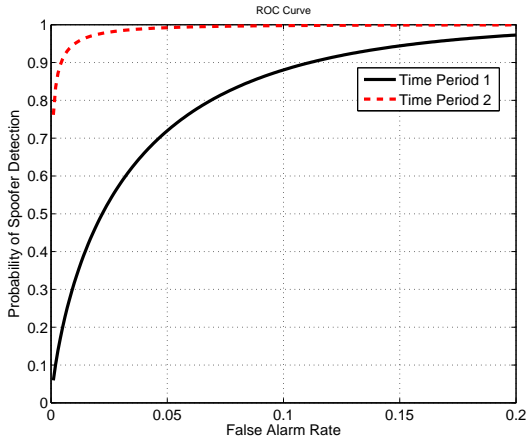}
  \caption{ROC curve for the $R_i$ test statistic in detecting a Spoofer}\label{fig:ROC}
\end{figure}

The PDF of the standard deviation of the $R_i$ values for each of the two time periods of the day, and the spoofing signal,
are shown in Figure \ref{fig:PDF}.
A non-central Chi distribution was fitted to the GPS signal PDF's and a chi distribution to the spoofer PDF. The Chi distribution was chosen as the
test statistic is the standard distribution of a set of random variables.  It is clear from Figure \ref{fig:PDF}
that the PDFs have a better separation during the second time period.
This is due to a more favorable satellite geometry during this time period.
This suggests that the detection threshold and false alarm rate should be adjusted based on the time of day.  The satellite orbits are predictable and
repeat daily, allowing the receiver to build up a PDF of the expected standard deviation of the $R_i$ test statistic for the GPS signal only (no spoofer)
 case for each time period of the day and hence determine the optimal threshold, false alarm rate and probability of detection for that particular period.
The receiver operation character (ROC) curves for the two $90$ minutes periods over which the data was logged are shown in Figure \ref{fig:ROC}
and clearly show a much better detection performance during the second time period indicating a more optimal satellite geometry.

In summary, the spoofing detection algorithm on the physical layer provides priori information of whether
PMU $n$ is suffering from TSA, which is given by the likelihood ratio:
\begin{equation}\label{Eq.LikeDete}
    \mathcal{L}_{PHY} = \frac{p(y_{PHY}(n))|T_n=H_1}{p(y_{PHY}(n))|T_n=H_0},
\end{equation}
where $p(y_{PHY}(n))=std([R_1,R_2,...,R_i...R_N])$;
$T_n=H_1$ and $T_n=H$ denote the hypotheses of whether PMU $n$ is under TSA or not, respectively.
The probability of $p(y_{PHY}(n))|T_n=H_1$ and $p(y_{PHY}(n))|T_n=H_0$ can be further used as the priori information for
the trustworthiness evaluation in the upper layer, which will be discussed in the subsequent sections.

\section{Trustworthiness Based TSA Detection}\label{sec: Trustiness}
In this section, we discuss the TSA detection in the upper layer, i.e., from the viewpoint of signal processing. We first introduce the linear system model for the power grid. Then, we propose a mechanism for evaluating the trustworthiness of each monitoring device based on
the linear system model.

\subsection{Linear System Model}
Generally, smart grid is a wide area interconnected nonlinear system.  However, for the purpose of designing
a WAMS in a smart grid, the dynamic feature of the entire power grid can be modeled as a linear system around the equilibrium point, given small perturbations \cite{LinearPSys_Liu}\cite{LinearPsys_Machowski}. We will focus on linear system model in this paper and extend to nonlinear case in the future.

Mathematically, the linearized power system can be expressed as a state space model, which is given by
\begin{eqnarray}
  \mathbf{x}(t+1) &=& \mathbf{A}\mathbf{x}(t) + \mathbf{B}\mathbf{u}(t)+\mathbf{w}(t) \\
  \mathbf{y}(t) &=& \mathbf{C}\mathbf{x}(t)+\mathbf{v}(t)
\end{eqnarray}
where $\mathbf{x}$ is an $N\times 1$ vector and represents the system operation state of the power grid, $\mathbf{y}$ is an $M\times 1$ vector representing
the monitoring measurements and $\mathbf{u}$ is the control action taken by the control center. The matrices $\mathbf{A}$, $\mathbf{B}$, and matrix $\mathbf{C}$
are specified in the linearized system model and the monitoring mechanism, respectively. Both $\mathbf{w}(t)$ and $\mathbf{v}(t)$
are modeled as Gaussian noise. Note that the dynamics in power grid are continuous in the time domain. Here we consider a discrete time model for simplicity. We further suppose that each dimension of the measurement vector
$\mathbf{y}$ is monitored by a monitoring device, e.g., PMU. It is easy to extend this to the general case in which the monitoring measurements have overlaps.
Each monitoring device sends its measurements along with their time stamps to the control center via a secure communication network since they are
distributed throughout the entire smart grid. For the TSA on the monitoring devices, we have following assumptions:
\begin{itemize}
  \item For simplicity, we assume that there is at most one attacker. The principle of the trustworthiness of the system can be extended to the case of
        multiple attackers at the cost of more computational cost.
  \item The malicious attacker cannot modify the monitoring measurements; it can only trigger the monitoring device to sample at an incorrect time period using GPS spoofing.
          Therefore, the time stamps attached to those measurements are incorrect.
  \item The control center misaligns the measurements from the monitoring devices under TSA, which is equivalent to shifting the measurements in the time domain.
\end{itemize}

We assume that the controller adopts the linear quadratic regulation (LQR) control \cite{PsysControl_Kwakernaak} in an infinite time horizon \cite{PSySControll_Wu} with
the cost function given by
\begin{equation}\label{Eq. CostFunc}
    J = E[\sum_{t=0}^{\inf}\beta(t)(\mathbf{x}^T(t)\mathbf{Q}\mathbf{x}(t)
                      +\mathbf{u}^T(t)\mathbf{P}\mathbf{u}(t))],
\end{equation}
where $\mathbf{Q}$ and $\mathbf{P}$ are both positive definite, and $\beta$ is a weighting factor for each
control time period. Based on the cost function in \ref{Eq. CostFunc}, the LQR action $\mathbf{u}(t)$ is given by
\begin{equation}\label{eq.ut}
    \mathbf{u}(t) = -(\mathbf{B}^T\mathbf{S}\mathbf{B}+\mathbf{P})^{-1}\mathbf{B}^T\mathbf{S}\mathbf{A}\hat{\mathbf{x}}(t),
\end{equation}
where $\hat{\mathbf{x}}(t)$ is the estimation of the system state, and the matrix $\mathbf{S}$ satisfies the Algebraic Riccati Equation, which
is given by
\begin{equation}\label{Eq.S}
    \mathbf{S}=\mathbf{A}^T(\mathbf{S}-\mathbf{S}\mathbf{B}(\mathbf{B}^T\mathbf{S}\mathbf{B}+\mathbf{P})^{-1}\mathbf{B}^T\mathbf{S})\mathbf{A}+\mathbf{Q}.
\end{equation}

\subsection{Trustworthiness Evaluation}
\begin{figure}[htcp]
  \centering
  \includegraphics[scale=0.5]{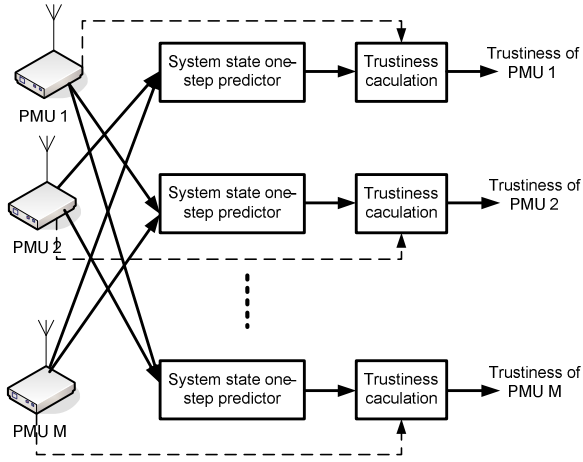}
  \caption{Illustration of trustworthiness evaluation}\label{Fig:TrustEv}
\end{figure}

In this subsection, we propose a mechanism for evaluating
the trustworthiness of each monitoring device. The essential reason that the
control center can evaluate the trustworthiness of each monitoring device is that the
observations at different PMUs are coupled and thus provide redundancies (similarly to error correction codes in communications). Therefore,
it can predict the future state with some uncertainty.
If the report from a monitoring device significantly deviates from the
prediction, then this monitoring device will be considered as
unreliable and its reports will be ignored.
Figure \ref{Fig:TrustEv} illustrates the mechanism of trustworthiness evaluation,
in which the monitoring device is a PMU that monitors the
frequency of the power grid.

The system state prediction is obtained from the Kalman filter \cite{Kalman_poor}.
At time $t+1$, we apply the system state prediction $\hat{\mathbf{x}}(t)$  and
the measurements $\mathbf{y}_{-n}(t+1)$ to predict the
system state $\hat{\mathbf{x}}(t+1)$, in which $\mathbf{y}_{-n}(t+1)$ denotes the measurements excluding
the one from PMU $n$. The Kalman predictor is given by
\begin{equation}\label{Eq.Kalman}
    \hat{\mathbf{x}}(t+1) = \mathbf{A}\hat{\mathbf{x}}(t)+\mathbf{L}_{-n}(t+1)[\mathbf{y}_{-n}(t)-\mathbf{C}_{-n}\mathbf{A}\hat{\mathbf{x}}(t)],
\end{equation}
where $\mathbf{C}_{-n}$ denotes the matrix $\mathbf{C}$ excluding the $n$-th row. The covariance matrix $\mathbf{L}_{-n}$ is given by
\begin{equation}\label{Eq.L}
    \mathbf{L}_{-n}(t+1) = \mathbf{\Sigma}(t+1|t) \mathbf{C}_{-n}^T\left[\mathbf{C}_{-n}\mathbf{\Sigma}(t+1|t)\mathbf{C}_{-n}^T+\mathbf{\sigma}^2_n\mathbf{I}\right]^{-1},
\end{equation}
where $\mathbf{\Sigma}$ is the prediction covariance, which is given by
\begin{equation}\label{Eq.Sigma}
    \mathbf{\Sigma}(t+1|t) = \mathbf{A}\mathbf{\Sigma}(t|t)A^T+BQB^T
\end{equation}
where $\sigma^2$ is the noise variance in the observations.
After we obtain the system state prediction $\hat{\mathbf{x}}(t+1)$, the a posteriori probability of the measurement from
PMU $n$ is given by
\begin{eqnarray}
  &&p(y_n(t+1)|\mathbf{y}_{-n}(0:t))  \nonumber\\
  &\sim & \mathcal{N}(\mathbf{c}_n\hat{\mathbf{x}}(t+1),\mathbf{c}_n\mathbf{\Sigma}(t+1|t)\mathbf{c}_n^T),
\end{eqnarray}
where $\mathbf{c}_n$ denotes the $n$-th row of the observation matrix $\mathbf{C}$.
We denote by $T_n=H_1$ the hypothesis that PMU $n$ is an unreliable monitoring device due to TSA,
and $T_n=H_0$ as the hypothesis that the measurement from PMU $n$ is trustworthy.
Then, we define the trustworthiness level of PMU $n$ at time slot $t$ as $1-\pi_n(t)$, where
$\pi_n(t)$ is the suspicious level, which is given by
\begin{equation}\label{Eq.pi}
    \pi_n(t) \triangleq p(T_n=H_1|\mathbf{y}(0:t)).
\end{equation}
In the next section, we will derive the trustworthiness evaluation based on the system state prediction and the information passed from the physical layer as prior information, which is coined cross layer TSA detection. Hence, we ignore the details of the Kalman filtering based trustworthiness evaluation since it is only a special case of the cross layer detection.

\section{Cross Layer TSA Detection}\label{sec: CrossLayer}
In this section, we present the cross layer TSA detection algorithm.
Since both the spoofing detection mechanisms in the physical layer  (based on the two antennas) and the upper layer (based on the Kalman filtering)
are based on the probabilistic framework, we can integrate the detection schemes in the two layers to detect TSA. Based on the probabilistic framework,
the spoofing detection result from physical layer can be regarded as the prior
information for the upper layer to evaluate the suspicious level.
The challenge for computing the suspicious level defined by (\ref{Eq.pi}) is the unknown attacking strategy. We first assume that there must be an attacker.
With the physical layer's detection as the prior information, we can derive (\ref{Eq.pi}), which is given by
\begin{equation}\label{Eq,Pi2}
    \pi_n(t) \triangleq p(T_n=H_1|\mathbf{y}(0:t),\mathbf{y}_{PHY}^n(t)),
\end{equation}
where $\mathbf{y}_{PHY}^n(t)$ is the output of the physical layer detection on time slot $t$ from PMU $n$. Applying the Bayes rule upon (\ref{Eq.pi}), we have
\begin{eqnarray}\label{Eq.Bayesian}
   & &\pi_n(t) \nonumber\\
   &=&\frac{p(\mathbf{y}(0:t),\mathbf{y}_{PHY}^n(t)|T_n=H_1)}
               {p(\mathbf{y}(0:t),\mathbf{y}_{PHY}^n(t))} \nonumber\\
   & & \times\frac{p(T_n=H_1)}{p(\mathbf{y}(0:t),\mathbf{y}_{PHY}^n(t))}.
\end{eqnarray}
We assume that the outputs of the physical layer's detection and the upper layer's observation of PMU are independent of each other, thus resulting in
\begin{eqnarray}\label{Eq.Bayesian3}
   & &\pi_n(t) \nonumber\\
   &=&\frac{p(\mathbf{y}(0:t)|T_n=H_1)p(T_n=H_1)}
               {p(\mathbf{y}(0:t))p(\mathbf{y}_{PHY}^n(t))} \nonumber\\
   &&\times\frac{p(\mathbf{y}_{PHY}^n(t)|T_n=H_1)}{p(\mathbf{y}(0:t))p(\mathbf{y}_{PHY}^n(t))}\nonumber\\
   &=&\frac{p(\mathbf{y}(0:t)|T_n=H_1)}
           {\sum_{m=1}^N p(\mathbf{y}(0:t)|T_m=H_1)}\\
   &&\times\frac{p(\mathbf{y}_{PHY}^n(t)|T_n=H_1)}
   {(p(\mathbf{y}_{PHY}^n(t)|T_n=H_1)+p(\mathbf{y}_{PHY}^n(t)|T_n=H_0))}.\nonumber
\end{eqnarray}
The conditional distribution of $p(\mathbf{y}_{PHY}^n(t)|T_n=H_1)$ and $p(\mathbf{y}_{PHY}^n(t)|T_n=H_0))$ are given by the curves in Figure \ref{fig:PDF}.
Since we do not have the prior information about the attacker's TSA strategies,
it is reasonable to assume that the PMU's report under TSA in each time slot is independent.
Therefore, the first part of (\ref{Eq.Bayesian3}) can be decomposed into
\begin{eqnarray}\label{Eq.Decomp}
  && \frac{p(\mathbf{y}(0:t)|T_n=H_1)}
            {\sum_{m=1}^N p(\mathbf{y}(0:t)|T_m=H_1)}\nonumber \\
   &=&\frac{1}{1+\sum_{m\neq n}\frac{\prod_{s=0}^{t}p(\mathbf{y}(s)|T_m=H_1)}
              {\prod_{s=0}^{t}p(\mathbf{y}(s)|T_n=H_1)}}.
\end{eqnarray}
Since there is only one TSA attacker,
the unknown TSA strategies also make the observations at different
PMUs independent, which provides the following approximation:
\begin{eqnarray}\label{Eq.Simplify2}
    &&p(\mathbf{y}(s)|T_n=H_1)\\
    &\approx& p(y_n(s)|T_n=H_1)\prod_{k\neq n}p(y_k(s)|T_k=H_0).\nonumber
\end{eqnarray}
We further assume that $p(y_n(s)|T_n=H_1)$ is a constant since we have no knowledge about the attacker's strategy. Substituting (\ref{Eq.Simplify2}) into (\ref{Eq.Bayesian3}), we obtain
\begin{equation}\label{Eq.Bayesian4}
   \pi_n(t)=\frac{\prod_{s=0}^{t}\frac{1}{p(y_n(S)|T_n=H_0)}}
                 {\sum_{m=1}^N\prod_{s=0}^{t}{1\over p(y_m(S)|T_m=H_0)}}\eta_{PHY}^n(t),
\end{equation}
where
\begin{eqnarray}\label{Eq.Eta}
    &&\eta_{PHY}^n(t)\\
    &=&\frac{p(\mathbf{y}_{PHY}^n(t)|T_n=H_1)}
   {(p(\mathbf{y}_{PHY}^n(t)|T_n=H_1)+p(\mathbf{y}_{PHY}^n(t)|T_n=H_0))},\nonumber
\end{eqnarray}
where $\eta_{PHY}^n(t)$ is the physical layer's prior probability calculated using the curves in
Figure \ref{fig:PDF} directly.

When it is possible that there is no attacker, the first part of (\ref{Eq.Bayesian3}) can be modified as
\begin{eqnarray}\label{Eq.Decomp2}
  && \frac{p(\mathbf{y}(0:t)|T_n=H_1)p(T_n=H_1)}
            {p(\mathbf{y}(0:t),H_3)+p(\mathbf{y}(0:t),H_4)} \\
   &=&\frac{p(\mathbf{y}_{PHY}^n(t)|T_n=H_1)}
   {p(H_3)/p(H_4)+\sum_{m\neq n}\frac{\prod_{s=0}^{t}p(\mathbf{y}(s)|T_m=H_1)}
              {\prod_{s=0}^{t}p(\mathbf{y}(s)|T_n=H_1)}},\nonumber
\end{eqnarray}
where $H_3$ and $H_4$ denote the hypotheses that there is no attacker and that there is one attacker, respectively.

\section{Experiments Results}\label{sec: Experiment}
In this section, we conduct experiments to demonstrate the proposed cross layer TSA detection algorithm. The TSA detection reports in the physical layer are obtained from the experiment
setup in subsection \ref{subsec:Experiment}. The results of the physical layer detection will be fed to the upper layer for trustworthiness evaluation.

\subsection{Upper Layer Linear System model}
We adopt the linear model of power grid used in \cite{LinearPSys_Liu}, in which the system state matrix $\mathbf{A}$ is given by Eq. (13) in \cite{LinearPSys_Liu}.
Since we discuss the discrete-time model in this paper, we approximate the continuous-time state space model by setting a small time step $\Delta t$. Therefore, the discrete-time state space equation is given by
\begin{equation}\label{Eq.State}
    \mathbf{x}(t+1) = (\mathbf{I}-\Delta t\mathbf{A})\mathbf{x}(t)+\Delta t\mathbf{B}\mathbf{u}(t).
\end{equation}
There are five PMUs in the system, and we only consider the frequency measurement.
Consequently, we set the observation matrix as a $5\times 5$ identity matrix.
We further assume that PMU $5$ encounters TSA and
other PMUs operate normally.

\subsection{Upper Layer Evolution of Suspicious level}
\begin{figure}[]
\vspace{0pt}
\subfigure[Constant time label shift TSA\label{Fig:Fixmove}]{
\begin{minipage}[b]{1\linewidth}
\centering
\includegraphics[scale=0.4]{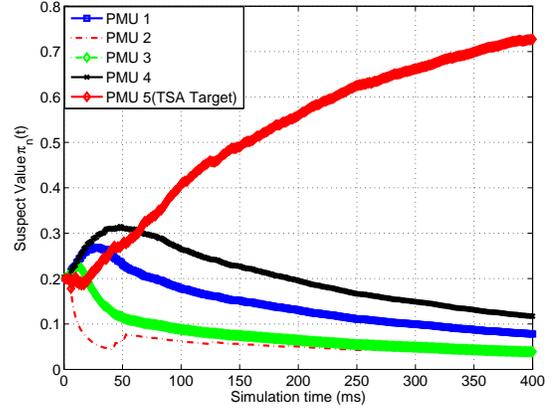}
\end{minipage}}\\%
\subfigure[Random time label shift TSA\label{Fig:RandomMove}]{
\begin{minipage}[b]{1\linewidth}
\centering
\includegraphics[scale=0.5]{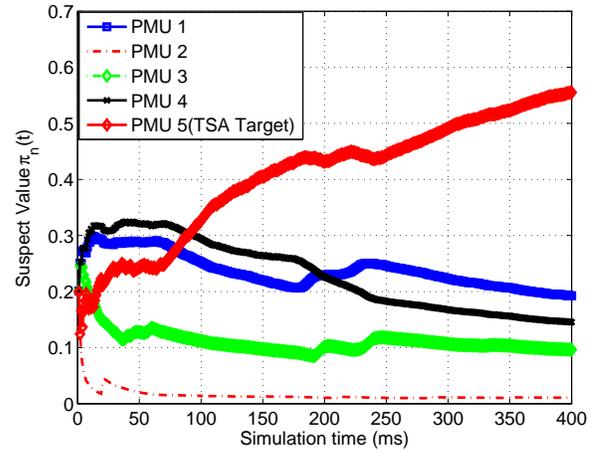}
\end{minipage}}
\caption{Suspicious level of different PMUs under different TSA attack strategies} \label{Fig:TrustCruve}
\end{figure}

In Figure \ref{Fig:TrustCruve}, we demonstrate the evolution of
the suspicious level when the attacker adopts different attack strategies.
The attacker launches TSA on PMU $5$ with attack frequency of $0.3$,
which means the attacker at each time slot launches TSA with probability
of $0.3$.
In Figure \ref{Fig:Fixmove}, the attacker modifies the time stamps with
a constant shift value, and in Figure \ref{Fig:RandomMove}, the attacker modifies
the time stamps with a random shift value.
The simulations show that the suspicious level of PMU $5$ increases
significantly after some fluctuation regardless of
the strategies the attacker applies.
The fluctuation in the initial stages are due to the randomness in
the Kalman filtering, since it takes time for the Kalman filter to
track the system state.

\subsection{Cross Layer TSA Detection}
\begin{figure}[htcp]
  \centering
  \includegraphics[scale=0.4]{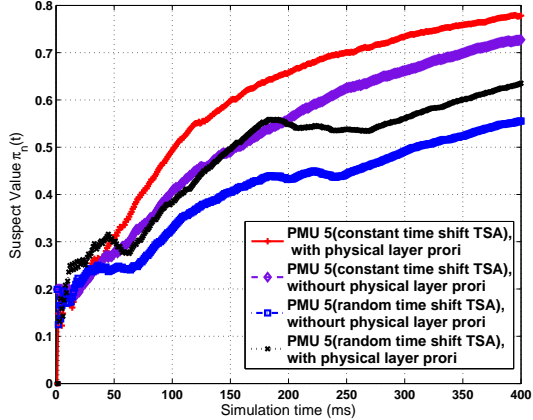}
  \caption{The performance comparison of TSA detection with physical layer prior and without physical layer prior}\label{Fig:TrustCross}
\end{figure}

Then, we simulate the proposed cross layer TSA detection algorithm.
Figure \ref{Fig:TrustCross} demonstrates that the suspicious level of the
PMU under TSA increases faster when the physical layer's detection reports
are used as the prior information.

\begin{figure}[htcp]
  \centering
  \includegraphics[scale=0.5]{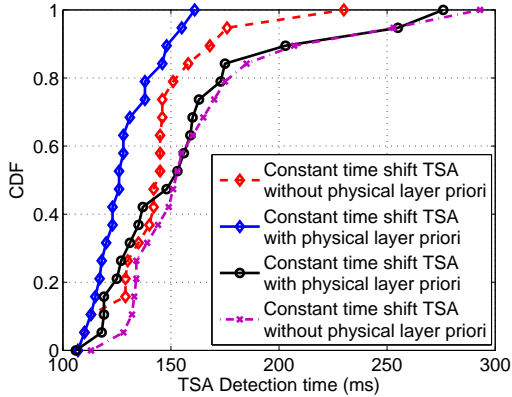}
  \caption{CDF curves of the time of identification of TSA}\label{Fig:CDF}
\end{figure}

In Figure \ref{Fig:CDF}, we plot the cumulative distribution function (CDF) curves of the time needed for detecting the TSA. This further verifies that
cross layer TSA detection can
identify TSA faster than the situation when only upper layer trustworthiness evaluation is used.
It should be noted that the detection performance is improved regardless the attacker's
strategies.

\begin{figure}[htcp]
  \centering
  \includegraphics[scale=0.45]{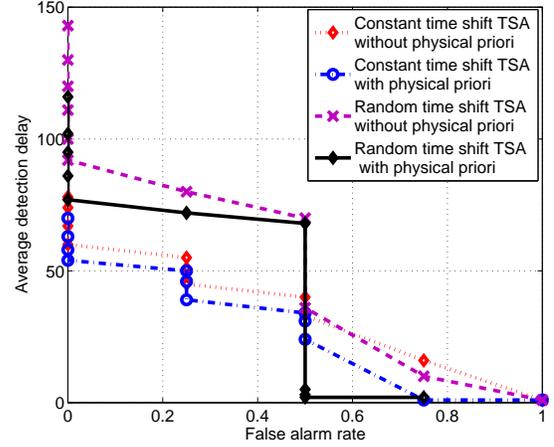}
  \caption{ROC curves (detection delay and false alarm) for different attack strategies}\label{Fig:ROCDelay}
\end{figure}

The receiver operation character (ROC) curves are given by Figure \ref{Fig:ROCDelay}.
In contrast to typical ROC curves, we study the average TSA detection delay and the false alarm rate. The false alarm rate is defined as the event that a PMU not suffering from TSA is claimed to be under TSA.
It is observed from Figure \ref{Fig:ROCDelay} that, given the same detection delay, the proposed cross layer TSA detection has lower false alarm rate compared with the TSA detection without the collaboration between the physical layer and upper layer.

\section{Conclusion}\label{sec:Conclusion}
In this paper, we have proposed a cross layer detection mechanism to combat time synchronization attack in smart grid. In the physical layer, we apply patch-monopole hybrid antenna to receive GPS signal, which will be fed to two GPS receivers. The difference of the C/No from the patch and monopole is used to estimate the probability of being under TSA. The experiment has shown that
the standard derivation of the difference of the C/No from two GPS receivers follows different distributions.
In the upper layer, we have applied the Kalman filtering and cross check to evaluate the trustworthiness of the reports. Furthermore we have fused the TSA detection result in the physical layer, as prior information, with the upper layer detection. Numerical results have demonstrated that the cross layer detection scheme can
effectively improve performance, with faster detection speed or lower false alarm rate.



%
%
%
%
%
%
%
%
%
%
\end{document}